\documentclass{aastex}
\usepackage{spr-astr-addons}
\usepackage{url}

\RequirePackage{color}

\begin{document}

\title{A unified model of supernova driven by magnetic monopoles} \slugcomment{Not to appear in
Nonlearned J., 45.}
\shorttitle{A unified model of supernova} \shortauthors{Qiu-He Peng
et al.}

\author{Qiu-He Peng \altaffilmark{1}, Jing-Jing, Liu \altaffilmark{2}, and Chih-Kang Chou \altaffilmark{3}}
\affil{Corresponding to Jing-Jing, Liu. (liujingjing68@126.com)}
\email{qhpeng@nju.edu.cn, and liujingjing68@126.com}

\altaffiltext{1}{Department of Astronomy, Nanjing University,
Nanjing, Jiangshu 210000, China.} \altaffiltext{2}{College of Marine
Science and Technology, Hainan Tropical Ocean University, Sanya,
Hainan 572022, China.}
\altaffiltext{3}{National Astronomical
Observatory, Chinese Academy of Sciences, Beijing, 100000, China.}

\begin{abstract}
In this paper, we first discuss a series of important but puzzling
physical mechanisms concerning the energy source, various kinds of
core collapsed supernovae explosion mechanisms during central
gravitational collapse in astrophysics. We also discuss the puzzle
of possible association of $\gamma$-ray burst with gravitational
wave perturbation, the heat source for the molten interior of the
core of the earth and finally the puzzling problem of the cooling of
white dwarfs. We then make use of the estimations for the space flux
of magnetic monopoles (hereafter MMs) and nucleon decay induced by
MMs (called the Rubakov-Callen(RC) effect) to obtain the luminosity
due to the RC effect. In terms of the formula for this RC
luminosity, we present a unified treatment for the heat source of
the Earth's core, the energy source for the white dwarf interior,
various kinds of core collapsed supernovae (Type II Supernova
(SNII), Type Ib Supernova (SNIb), Type Ic Supernova (SNIc), Super
luminous supernova (SLSN)), and the production mechanism for
$\gamma$-ray burst. This unified model can also be used to
reasonably explain the possible association of the short
$\gamma$-ray burst detected by the Fermi $\gamma$-ray Burst
Monitoring Satellite (GBM) with the  LIGO gravitational wave event
GW150914 in September 2015.
\end{abstract}

\keywords{stars: supernovae, stars: evolution, Physical Date and
Processes: nuclear reactions.}


\section{The puzzle of the explosion mechanisms for the various
kinds of core collapsed supernovae}

Although the true reason of supernovae explosion due to
gravitational collapse of the central core is not thoroughly
understood so far, considerable insight has been obtained concerning
both the physics of the gravitational collapse of the whole star
after the end of thermal nuclear fusion for massive stars, and the
initial stage of the collapse and the physical scenario why the
entire star can not explode \citep{Bethe90, Woosley86, Woosley91}.
The masses of the pre-main sequence supernova progenitor due to
gravitational collapse of the central core may be roughly classified
as follows: SNII, SN1b, SN1c (the physical origin for the long
$\gamma$-ray burst) with messes respectively given by
$M=(8-25)M_\odot$, $M=(30-60)M_\odot$  and $M=(80-150)M_\odot$ . In
addition, we also estimate the pre-main sequence mass for the
brightest supernovae SLSN to be $M=(200-1000)M_\odot$ or more.

For the supernovae SNII and SNIb , the physical mechanism for the
gravitational collapse of the stellar core is as follows. An iron
core with high temperature $T_c\approx (3-5)\times10^9$K and high
density $\rho_c\approx (2-3)\times10^9 \rm{g/cm^3}$ is formed after
the end of thermal nuclear fusion in the central core of massive
stars. The election gas is in a highly relativistic degenerate state
such that the Fermi energy of the election gas is obviously higher
than the energy threshold for electron captured by iron nuclei,
thereby causing large amount of the electrons quickly captured by
the iron nuclei (and by the elements of the iron family) via the
electron capture processes. Since large amount of free electrons
quickly break into the nucleus the resulting electron pressure in
the core is greatly reduced, and it then causes the entire iron core
to a fast gravitational collapsing. In the outer region the speed of
the collapse towards the center almost reaches half that of the free
fall. In the inner region, however, it was generally believed that
the core collapse was homogeneous as first suggested by
\citet{Colgate66}. But in 2004, \citet{Peng04} pointed out that
since the electron capture process increases very fast as the
density is increased, the resulting electron pressure decreases even
faster with the radial distance towards the center. Consequently,
the gravitational collapse in the central region should be in an
accelerating state. Our recent research series showed that the weak
interaction (e.g., electron capture and beta decay) plays a critical
role in the process of supernova explosion (e.g., \citet{Liu13,
Liu14, Liu16, Liu16a, Liu17}).

General idea of the collapsed process of the core collapse of the
supernova is following. The collapsing speed of matter at the
interface of the inner and outer regions may reach as high as
$(1/8-1/4)$c, where c is the speed of light in vacuum. The central
density of the star reaches $\rho_c\approx (2-5)\rho_{nuc}$ (where
$\rho_{nuc}$ is the nuclear density), the supporting pressure
against collapse becomes the non-relativistic degenerate neutron gas
which is both dynamically and thermodynamically very stable (the
non-relativistic degenerate neutron pressure is much higher than the
degenerate pressure of the relativistic electron gas). The collapse
of the inner region is quickly stopped.  The non-relativistic and
degenerate neutron gas not only resists the inward collapse of the
matter with lower density in the outer region but also drive the
matter with lower density in the outer region moving violently
outwards with tremendous speeds and a so-called rebound shock with
Mach number $(1-2)$ is then formed outside the interface between the
inner and outer region. The rebound shock wave carries with it huge
amount of energy (i.e., his energy is transformed from the
gravitational energy released by the matter in the core during the
collapse, it may reach $10^{52}-10^{53}$ ergs). The temperature of
rebound shock front may reach as high as $10^{11}$K. At such
temperatures, large number of the thermal $\gamma$-ray photons may
have energy about 10Mev and these energetic thermal$\gamma$-ray
photons immediately break the iron nucleus (and the elements of iron
family) into nucleons and ¦Á particles:
$^{56}\rm{Fe}+\gamma\rightarrow 13\alpha+4n$,
$\alpha+\gamma\rightarrow2p+2n$.

Since this process consumes huge amount of energy and all reliable
stellar evolution models predict a rather large mass for the outer
region of the iron core thereby causing the energy of the rebound
shock wave completely depleted before it can break and completely
destroy the outer region of the collapsing star. In other words, it
is impossible to make the whole outer region of the star to explode
outwards and this "instantaneous explosion mechanism" for supernovae
explosion fails. If we adopt the idea the of accelerated collapse of
the central core as proposed by \citet{Peng04} which is due to the
increase of the electron capture rates with density at the
supernovae core, the mass of the inner core during the initial stage
of the collapse clearly decreases. Along this line of thought, we
also make use of the simplest model for gravitational collapse with
spherical symmetry and some artificial parameter to simulate the
collapse, The supernovae explosion is possible for some of our
models with appropriate parameters \citep{Luo08}.

Because the failure of the instantaneous explosion mechanism Wilson
proposed a neutrino delayed mechanism for supernovae explosion
\citep{Wilson88}. After the collapse of the supernovae core a
nascent neutron star with high temperature $10^{11}$K is formed.
Wilson proposed that a strong neutrino flux will be produced from
the nascent neutron star. The weak interaction between the strong
neutrino flux with the matter of outer region of the supernovae can
lead to explosion of the outer region and forming neutrino
explosion. However, two questions are open for the neutrino delayed
mechanism: a) How this strong neutrino flux is generated?  b)
Whether the supernova can be expelled by the weak interaction
between the strong neutrino flux with matter in the outer region of
the supernova?

Based on high temperature, $10^{11}$K, of a nascent neutron star
from the collapse of the supernovae core, we have proposed that in
the high temperature interiors of the nascent neutron stars by
supernovae core collapse, large amounts of neutrino flux with energy
10Mev maybe produced in less than $10^{-6}$ second via the phase
transition processes (i.e. $u+e\rightarrow d+\nu_e$, $u+e\rightarrow
s+\nu_e$, $u+d\rightarrow u+s$) from the two flavor quarks (u, d)
into the there flavor quarks (s, u, d) with total energy about
$10^{52}$ erg \citep{Dai95}. Nevertheless, it is still open question
up to now whether the interaction between strong neutrino flux with
matter can really generate the strong outward pressure to achieve
supernovae explosion with outward initial speed $10^4$km/s and total
kinetic energy $10^{49}$ergs? Although considerable efforts have
been devoted to study neutrino transport process and neutrino fluid
dynamics with complicate (2D and 3D) numerical simulation,
unfortunately supernovae explosion through self consistent theory
and simulation has not been achieved. The representative reference
in this regard may be illustrated by \citet{Buras03}, and
\citet{Lie04}.

It is generally believed that the progenitor of Ic type supernovae
(SNIc) is the massive pre-main sequence with large mass
$M=(80-150)M_\odot$ and the progenitor
 of the brightest supernovae (SLSN) is the pre-main sequence star with even large mass
($M>(200-500)M_\odot$ ). These supermassive stars have already,
lived through the stages of hydrogen burning (main sequence star)
and helium burning (red giant star). Since the reaction rate of
$^{12}\rm{C}+\alpha\rightarrow ^{16}\rm{O}+\gamma$ is very fast, the
abundance of $^{12}$C in the core after helium burning may be lower
than 8\%. The small amount of $^{12}$C may continue to burn during
the core collapse. The central region of the star then start to burn
oxygen directly and the temperature of the core may reach
$5\times10^9-10^{10}$K. At the high temperature region, a large
number of high energy thermal $\gamma$-ray photons annihilate into
election position pairs and the inverse reaction
($\gamma+\gamma\rightarrow e^++e^-$) may reach thermal equilibrium.
Meanwhile, large number of the selection positron pairs may
annihilate into neutrino pairs at very fast rate and these energy
outgoing neutrinos immediately escape from the star.

The energy carried away by the neutrinos is lost from the star and
lead to the fast reduction of the interior thermal pressure and
finally resulting in drastic gravitational collapse of the
supernovae lore. This is well known as the instability generated by
electron positron pair annihilation into neutrino pairs as proposed
by \citet{Chiu60}. This is the physical mechanism for the
gravitational collapse of super-massive stars. From some
astrophysical observational evidence it is generally that long
$\gamma$-ray burst is associated with type Ic supernovae explosion.
The explosion mechanism for SNIc has been investigated in terms
numerical simulation since 2007. No critical progress has been
achieved. Thus, the explosion mechanism for SNIc is still a puzzle.

As for the super luminous supernova (SLSN), the absolute magnitude
of their peak luminosity is $M_{\rm{peak}}\leq -21^m$ (but the peak
luminosity of the supernova usually is $M_{\rm{peak}}\geq-20^m$).
The ultra luminous supernova ASASSN-15lh discovered in Sep. 2015
\citep{Dong16}, after 15 days from the peak luminosity, its
luminosity is still up to $2.2\times10^{45}\rm{erg/s}$. It is twice
as bright as the brightest supernova luminosity up to date. Its
radiation energy is up to $1.2\times10^{45}$ergs during four months.
Recently, the discussion about it is quite warm. For example,
\citet{Dai16} roughly guess the ultra bright supernova ASASSN-15lh
might being the evidence for the existence of quark stars. However,
this paper does not involve the discussion of the mechanism of the
supernova explosion. The ASASSN-15lh phenomenon is still a doubt. In
short, the mechanism of various types of supernova explosion is
still a major mystery.

The plan of this paper is as follows. In Section 2, we introduce a
new insight into MMs and their properties. In Section 3, we discuss
astrophysical evidence for the existence of MMs. We introduce our
model of quasars and AGNs containing MMs proposed in 1985
\citep{Peng85a, Peng85b} and predictions of the model proposed in
\citet{Peng01}. It seems that the subsequent observational evidence
favorably support our early predictions and the observed
astrophysical evidences support the existence of MMs. We detailed
discuss the models of the compelling mechanism for the various types
of supernovae explosion in Section 4. In section 5, we investigate
the other related puzzle of our model of supernova explosion driven
by magnetic monopoles. For instance, the energy source for the hot
molten interior of the earth's core and the puzzle of cooling of the
white dwarfs are discussed. The collision and merge of two massive
neutron stars containing MMs are also used to explain the possible
association of the short $\gamma$-ray burst with the LIGO
gravitational wave perturbation event GW150914 in Section 5. Finally
some conclusions and outlooks are given in Section 6.


\section{Study on the Magnetic Monopole}
\subsection{Rubakov-Callen(RC) effect about Nucleon Decay induced by MM}

During the period between 1970 and 1980 of the last century,
particle physicists are especially interested in discussing the
problem of MMs \citep{Polyakov74, Ma83, Shnir05, 't Hooft74}. It is
generally believed that super-massive MMs probably exist. In that
times its mass was estimated as $m_m\sim 10^{16} m_p$, where $m_p$
is the mass of the proton. But it is estimated as $m_m
\sim(10^3-10^4 )m_p$ in recent years \citep{Aab16}.

The magnetic charge of a MM is
$g_m=3hc/(2e)=0.988\times10^{-7}$(c.g.s). As for possible
applications to astrophysics, the most important property of MMs is
that these exotic particles can induce nucleon decay, namely
\begin{equation}
\label{eq.1}
  pM\rightarrow e^+\pi^0M+\rm{debris}~~~~~~(85\%),
\end{equation}
\begin{equation}
\label{eq.2}
  pM\rightarrow e^+\mu^{\pm}M+\rm{debris}~~~~~~(15\%),
\end{equation}
where pM means that a proton ( p) catalyzed by magnetic monopoles
(M) decays into a positron and a neutral pi meson ($\pi^0$) or
decays into a positron and a pair of $\mu^{\pm}$.

These MM induced reactions are proposed independently by
\citet{Rubakov81} and \citet{Callan83} with reaction cross section
$\sigma=(10^{-25}-10^{-26})\rm{cm}^2$. On the other hand, a much
lower cross section $\sigma\approx(10^{-36})\rm{cm}^2$ is proposed
by \citet{Wilczek82}. For simplicity, the reaction of MM induced
nucleon decay is called the RC effect with the estimated typical
cross section $\sigma\approx(10^{-30})\rm{cm}^2$ or
$\sigma\approx(10^{-30\pm6})\rm{cm}^2$.

\subsection{The amount of MMs in space}

It is generally believed that during the phase transitions in the
very early universe, small amounts of MMs may be generated by the
drastic oscillation and thermal agitation of the Higgs field . The
amount of MMs may by defined as the ratio of the number of MMs,
$N_m$, to that of the baryons, namely, $\xi=N_{\rm{M}}/N_{\rm{B}}$,
where $N_B$ is the number of baryons. Here we define a parameter
$\zeta_s$ ,named Newton saturation content, when the Newtonian
gravity of a MM from the center of the stars, balance with the
magnetic Coulomb repulsion by the interior total magnetic charge
(with the same polarity). Newton saturation content is the ratio of
the maximum number of MMs at the center of the stars and the number
of baryons in accumulation area of the MM. Thus, the Newton
saturation value is \citep{Peng85a}
\begin{equation}
\label{eq.3}
\zeta_s=Gm_{\rm{B}}m_m/g_m^2\approx1.9\times10^{-32}\frac{m_m}{10^9m_p},
\end{equation}
where we take $10^9 m_p$ as the mass of the magnetic monopole, which
is intermediate between $10^4 m_p$ \citep{Aab16} and $10^{16} m_p$
\citep{Polyakov74, 't Hooft74}.

\subsection{The amounts of MMs contained in astronomical bodies}

The following discussions and predictions have been presented in our
paper in 1985 \citep{Peng85a, Peng85b}

Firstly, in the early universe matters were in the ionized plasma
state with very high temperature. The strong electromagnetic
interaction between the MMs with plasma may lead to the existence of
considerable amount of the MMs in the central region of astronomical
bodies. This is because the formation of quasars and active galactic
nuclei (AGNS) from huge and super-massive primordial cloud through
gravitational collapse of the plasma via Jeans instability. Thus,
the central region of our Milky Way galaxy, quasars and ANGS may
contain more MMs with the Newtonian saturation value $\zeta_s$. in
section 3, we will discuss in more detail of our model for
super-massive stellar objects with MMs . This model can provide huge
amount of radiation and we may use it to replace the usual black
hole model as the energy source.

Secondly, generally speaking, stars and planets are formed in the
massive neutral hydrogen cloud via Jeans gravitational instability.
The interaction between MMs with neutral hydrogen atoms is very weak
so that there are very few of the MMs can fall towards the central
region following gravitational collapse of the neutral hydrogen
cloud and they are neglected. But some MMs may be contained in the
interiors of stars and planets now. These MMs are mainly captured
from space during their life time after their formation.  The flux
of the MMs captured from space may be estimated as follows \citep{
Peng85a}
\begin{eqnarray}
\label{eq.4}
&\phi_m=n_m\nu_m=10^{-4}c\zeta_m^0n_{\rm{B}}^0(\frac{\nu_m}{10^{-4}c})\nonumber\\
&=6\times10^{-12}(\frac{\zeta_m^0}{10^{-20}})(\frac{m_m}{10^9m_p})^{-1}(\frac{n_{\rm{B}}^0}{1\rm{cm}^{-3}})(\frac{\nu_m}{10^{-4}c})~~\rm{cm^{-2}s^{-1}},\nonumber\\
\end{eqnarray}
where the superscript (0) denotes for the space, and $n_{\rm{B}}^0$
denotes the number density of baryons in the interstellar space of
the Milky Way galaxy,  $\nu_m$ represents the velocity of the MMs in
space, and c is the speed of light. We use $\zeta_m^0$ to denote the
value $\zeta_m\equiv(N_m/N_B)$ of MMs contained in space according
to the upper limit given by \citet{Parker70}, $\xi_m^0\leq10^{-20}$
. The Newtonian saturation value, $\xi_s$, is lower than the upper
limit given by Parker by five order of magnitude
($\xi_m^0/\xi_s\leq10^{5}$).

The total number of MMs captured from space by stars or planets
after their formation may be estimated to be \citep{Peng85a}
\begin{eqnarray}
\label{eq.5}
N_m&=4\pi R^2\phi_mt=5.0\times10^{28}(\frac{\zeta_m^0}{10^{-20}})(\frac{n_{\rm{B}}^0}{1\rm{cm}^{-3}})(\frac{m_m}{10^9m_p})^{-1}\nonumber\\
&\times(\frac{\nu_m}{10^{-4}c})(\frac{R}{R_\odot})^2(\frac{t}{4.5\times10^9\rm{Yr}})~~\rm{cm^{-2}s^{-1}}
\end{eqnarray}
where R denotes the radius of the star and t is the life time of the
star. Thus, the MMs captured from interstellar space by stars and
planets during their life time is proportional to the surface area
and the life time of the stars and planets. The time scale of the
stars in the main sequence stage is about 90\% of the life time for
the stars. Due to hydrogen burning in the main sequence stage of the
stars the stars is relatively stable and the stellar radius changes
little during the main sequence stage. However, the stellar radius
during the red giant stage of the stars will increases up to more
than 10 times of the radius of the main sequence stars at least ,
although the time scale of the stars in the red giant stage is only
about 10\% of the life time for the stars. Thus, the total number of
MMs captured from space by the progenitor of supernova in its life
time is about 10 times of one calculated by the formula (4) for this
progenitor during its main sequence stage.

Since MMs are super heavy ($m_m>10^3m_p$ at least) so that the
captured MMs accumulate in the stellar core with  concentration
$n_m=\zeta_sn_B$ . The radius of the core
$r_c\approx(N_m/\xi_sn_{\rm{B}}^c)^{1/3}$, where $n_B^c$ represents
the average number density of the nucleons in the central core. The
total mass of the MMs accumulated in the central region becomes only
\begin{equation}
\label{eq.6}
 M_m=N_mm_m\approx10^{-8}N_m\frac{m_m}{10^{16}m_p}~~~~\rm{gm},
\end{equation}

Although this mass of the MMs accumulated in the central core is
only a very small fraction of the total mass of the star, it can
possibly provide huge luminosity (called RC luminosity) determined
by the central density of the stars via the RC effect (nucleon decay
induced by MMs)

\subsection{The Reason why MMs were not detected by Geophysical Laboratories}

Most physicists are very disappointed by the fact that MMs were not
detected in the geophysical laboratories. Recently, the book from
Wikimedia collected all the physics experiments designed for the
detection of MMs and pointed out that all physics experiments fail
to detect the existence of MMs  until now. They also give the upper
limit for the ratio of the number of MMs to that of the nucleons as
$10^{-29}$. All these experiments are carried out either on the
surface of the earth or in the earth's outer layer. Besides, in 1969
NASA sent the satellite explorer 35 (orbiting the moon) with GSFC
magnetometer. Their scientific goal is to measure the difference
between the magnetic north pole and the magnetic south pole of the
moon so as to measure the number of MMs contained in the moon.

According to the measurements of \citet{Schatten70}, and
\citet{Schatten83}, they gave the upper limit for the ratio of the
number of MMs to that of the nucleons as $6\times10^{-23}$. In order
to carry out the detection experiments of MMs in the geophysics
laboratories we may rewrite Equ. (3) as
\begin{equation}
\label{eq.7}
\phi_m\approx0.25(\frac{\zeta_m^0}{10^{-20}})(\frac{m_m}{10^9m_p})^{-1}(\frac{n_{\rm{B}}^0}{1\rm{cm}^{-3}})(\frac{\nu_m}{10^{-4}c})m^{-2}~~\rm{Yr}^{-1}
\end{equation}

The expression (7) implies that the flux of the MMs in flight in
interstellar space is very small. This means that the detection
probability of captured magnetic monopoles in flight is only about
one in 4 years even for large detector area $1m^2$. This is the
reason why MMs have not yet been detected by cosmic ray detector
experiments and geophysical laboratories. The number of MMs captured
by the earth after its formation (the age of the Earth is about 45
hundred million years) according to Equ.(5) is given by
\begin{equation}
\label{eq.8}
N_m\approx5.0\times10^{24}(\frac{\zeta_m^0}{10^{-20}})(\frac{m_m}{10^9m_p})^{-1}(\frac{n_{\rm{B}}^0}{1\rm{cm}^{-3}})(\frac{\nu_m}{10^{-4}c})
\end{equation}

These super-heavy MMs all accumulate in the central region of the
earth with radius less than 1 Km. There are no MM in the mantle, the
outer layer or the surface of the Earth. All experiments designed to
detect MMs in the outer layer of the Earth are useless.
Consequently, there are no MM  in the outer crust of the earth.

The MMs captured by the moon is far lower than that captured by the
earth because much larger surface area of the earth. The radius of
the moon is about $1/4$ that of the earth and the age of the moon is
about the same as that of the Earth. The number of MMs accumulated
in the core of the moon may then be estimated to be
\begin{equation}
\label{eq.9}
N_m\approx6.2\times10^{22}(\frac{\zeta_m^0}{10^{-20}})(\frac{m_m}{10^9m_p})^{-1}(\frac{n_{\rm{B}}^0}{1\rm{cm}^{-3}})(\frac{\nu_m}{10^{-4}c})
\end{equation}
where the mass of the moon is about $1/81$ that of the earth , i.e.
$7.4\times10^{25}$ gm, thus, the ratio of the number of MMs to that
of the nucleons for the moon is
\begin{equation}
\label{eq.10}
\zeta\approx1.4\times10^{-27}(\frac{\zeta_m^0}{10^{-20}})(\frac{m_m}{10^9m_p})^{-1}(\frac{n_{\rm{B}}^0}{1\rm{cm}^{-3}})(\frac{\nu_m}{10^{-4}c}).
\end{equation}

The upper limit of the number of MMs contained in the Moon by the
1983 moon orbiting satellite experiments is $6\times10^{-33}$. This
observed upper limit just gives a restrictive condition, which is
given by
\begin{equation}
\label{eq.11}
(\frac{\zeta_m^0}{10^{-20}})(\frac{m_m}{10^9m_p})^{-1}(\frac{n_{\rm{B}}^0}{1\rm{cm}^{-3}})(\frac{\nu_m}{10^{-4}c})\leq4\times10^{-6}.
\end{equation}

We anticipate that future satellite experiments orbiting the Earth
or the Moon with more accurate GSFC magnetometer with improve
measurement by ($3\sim4$) orders of magnitude, so that the possible
existence of MMs in the center of the Moon or the Earth may be
determined. At least, Eq.(11) may provide a very good constraint.

Although geophysical laboratories can not determine the possible
existence of MMs contained in the earth or the moon, most recent
discovery in astrophysical observations bring some hints for the
possible existence of MMs in the core of huge massive stellar object
(see Section 3). This is because strong radial magnetic field with
strength $>8$MG was discovered in the vicinity 0.12pc of the central
region of our Milky way Galaxy in 2013. The details of the
connection between the discovery of strong radial magnetic field and
the existence of magnetic monopoles will be elaborated in Section 3.
We would like to present a unified treatment for energy sources that
may explain the explosion mechanisms for the various types of
supernova explosions (including $\gamma$-ray bursts). We also may
explain the phenomenon of the molten core for the Earth, cooling
problem of white dwarfs (WDs). It will be shown that MMs located in
the central region of the stellar objects may play the key role in
our theory.

\subsection{The discussions of RC luminosity}
The total luminosity generated by the nucleon decay induced by MMs
in the central region of various stellar objects may the estimated
as follows

\begin{equation}
\label{eq.12}
 L_m\approx\frac{4\pi}{3}r_c^3n_mn_{\rm{B}}\langle\sigma\nu_{\rm{T}}\rangle
 m_{\rm{B}}c^2=N_mn_B\langle\sigma\nu_{\rm{T}}\rangle m_{\rm{B}}c^2,
\end{equation}
where $r_c$ is the radius of the stellar central region where MMs
accumulate, $n_m, n_B$ are respectively the number density of MMs
and nucleons, $N_m$ is the total number of MMs in the core of the
stellar object and it is also the number of captured MMs since the
birth of the stellar object. In the Equ.(12), $\sigma$ denotes the
reaction cross section for the RC effect,
$\sigma\approx(10^{-25}-10^{-26})\rm{cm}^2, and \nu=\nu_{\rm{T}}$
represents the thermal velocity of the nucleons relative to the MM .
This is because MMs are super-heavy and their thermal velocities can
be neglected. Consequently, we consider only the thermal velocity of
the nucleon with $\nu_{\rm{T}}=\sqrt{2kT/m_{\rm{B}}}$, where T
denotes the temperature and k is Boltzmann's constant.
$\langle\sigma\nu_{\rm{T}}\rangle$ depicts the thermal average.

In the RC process MMs induced nucleon decay, followed by nucleon
decay into $\pi^0$ meson, $\mu^\pm$ leptons and positions $e^+$,
$\mu^\pm$ and $\pi^0$ again decay into photons rand election
position pairs $e^\pm$. The positions then annihilate with the
elections to photons. The net effect is that the rest mass energy of
nucleons ( $ m_{\rm{B}}c^2$) entirely converted to radiation energy
with 100\% efficiency.($
1m_{\rm{B}}c^2\approx1\rm{GeV}\approx10^{-3}\rm{ergs}$).

\section{Astrophysical evidence for the existence of MMs}

Since 1985 we have seriously interested in studying the following
problem: if the idea of MMs and the RC effect in particle physics
are correct, what are their important effects on theoretical and
observational astrophysics? \citep{Peng85a, Peng85b, Wang85}

\subsection{Model of Quasars and AGN with MMs}

 It is generally believed that the super-massive central regions of
quasar and AGNs are formed from the gravitational collapse of
primordial super-massive stellar clouds during the early epochs of
the universe via Jeans instability.

During the early epoch of the universe the temperature is so high
that the primordial stellar clouds are in the plasma state. Since
the electromagnetic interaction of the MMs with the stellar plasma
clouds is very strong such that large amounts of the MMs fall and
accumulate in the central region of the stellar clouds following
gravitational collapse via Jeans instability. These super-heavy MMs
accumulate in the central region as much as possible until the
Newtonian saturation value ($\zeta_s$) is reached. Using the RC
effect as the energy source we proposed the models for quasars and
AGNs with MMs 32 years ago \citep{Peng85a, Wang85}. The idea of our
approach is as follows

First, we use MMs induced nucleon decay in to leptons as the main
energy source of quasars and AGNs to replace the black hole model
(the accretion flow may be used as a secondary source of energy)
Second, the gravitational effect of the super-massive star
containing MMs is similar to that of a black hole. The gravitational
effects is almost the same for these two models outside 50 AU from
the center.  The matter in the massive central core is relatively
rarefied. The radio waves and infrared waves observed from the
direction of the Galactic Center (hereafter the GC) in recent years
are probably originated from the inner region of the super-massive
star at the GC .

We note that super-massive stellar object with sufficient MMs have
neither black hole horizon nor central singularity. This is because
the reaction rate of the RC effect (nucleon decay into leptons
induced by MMs) is proportional to the square of the matter density
so that the central density cannot become infinite. Making use of
the RC effect in particle physics we can avoid the problem of the
central singularity due to the black hole of classical general
relativity. In this way the theory of the physics involved is self
consistent and harmonious.

\subsection{Main Predictions of our model and the observational test}

In 2001, we put forward five theoretical predictions for a model of
a supper massive object with MMs in the GC \citep{Peng01}. The main
ideas are as follows

Large amount of positions, for our first prediction, are produced at
the GC with production rates $6.4\times10^{42}e^+\rm{s}^{-1}$. It is
in agreement with the high energy astrophysical observations
($(3.4-6.3)\times10^{42}e^+\rm{s}^{-1}$) in 2003 by detecting the
very strong spectral lines of election - position annihilation
observed along the direction of the GC \citep{Kn03}.

After the publication of this observational result, some theoretical
models to explain both the 511keV and the GeV Gamma-rays from the
direction of the GC appeared. For instance, during the period
$2005-2006$, Wang et al. proposed that there may be millisecond
pulsars exist in the galactic center (e.g., \citet{Wang05, Wang06}).
\citet{Boehm04} considered a dark matter model. \citet{Casse04}
assumed that there may exist $\gamma$-ray bursts in the GC.
\citet{Cheng07} proposed that black hole explosion activity may be
used to explain the 511 keV radiation. Our predications were
published \citep{Wang85} and reemphasized in our paper
\citep{Peng01}. We would like to emphasized again that our
predication for the production of large amount of positrons from the
GC is quantitatively verified by the observations in 2003
\citep{Kn03}.

The second prediction is that some strong high energy radiation with
energy higher than 0.511Mev is simultaneously emitted from the huge
massive stars with MMs at the GC. Their integral total energy is
much higher than that for the election - position annihilation lines
and much higher than the thermal luminosity of the central stellar
object. This is also in agreement with observations.

The third prediction is that the MMs accumulated at the center of a
supermassive stellar object can generate powerful radial magnetic
field. The magnetic field strength is about $(20-100)$ Gauss at the
stellar surface of radius 50 a.u. \citep{Peng01}.  This prediction
is also in good agreement with the observation in 2013 \citep{
Eatough13}. Since the field strength of the radial magnetic field is
inversely proportional to the square of the radial distance so that
$B\approx(10-50)$mG at $r=0.12\rm{pc}\approx3.1\times10^{17}$cm.
This prediction is also in good agreement with the lower limit 8mG
determined from the observations in 2013 \citep{Eatough13}.

In 2001, we pointed out that the prediction for the presence of
radial magnetic field in the vicinity of the super-massive object at
the GC can be tested in the not distant future by astrophysical
observations \citep{Peng01}. If powerful radial magnetic field is
really confirmed by astrophysical observations, then our model is
unique in the sense that our model can naturally predict the
existence of strong radial magnetic field.

Fourthly, extreme high energy cosmic rays with energy
($10^{18}-10^{21}$)eV have been suggested to originate from
accelerations in the AGN core region, jet shear layers
\citep{Letessier11}. but if we assume that all super-massive
galactic nuclei of the AGNs within the range of 50Mpc from the Earth
contain saturated MMs (see the details in paper of \citet{Peng02}).
Of cause, there are many possibilities of the source of extreme high
energy cosmic rays with energy in theory. For example, the review by
\citet{Letessier11} mentions conventional models such as
accelerations in the AGN core region, jet shear layers etc., and
also several exotic models. Indeed, these models have their
respectively issues, but in general, they should be considered as
reasonable theoretical models.

Finally, we predicted the surface temperature of the super-massive
stellar object at the GC to be 123 K and the corresponding peak
value of the thermal radiation is roughly $10^{13}$Hz (at the
sub-millimeter range) \citep{Peng01}, and this is quite close to the
observed value of $10^{12}$Hz \citep{Falcke13}.

Based on the above predictions as well as the new discovery and the
latest progress made in recent years, especially the discovery of
the unusually strong radial magnetic field near the GC in 2013, we
(see \citet{Peng16a, Peng16b, Peng16c} pointed out that such
powerful radial magnetic field necessarily stop the plasma in the
accretion disk surrounding the GC from entering the inner core
consequently, the large amount of radiation (radio, infrared and
X-ray ) observed from the direction of the GC can not be generated
by the accreting material. From this we assert that the black hole
model of the GC of our Milk Way that has been prevalent for almost
half a century must not be real. Our model of the super-massive star
containing MMs for the GC is a useful alternative model that can
explain the observed radiation from the GC. Moreover, in the Refs.
\citet{Peng16a, Peng16b, Peng16c}, we have shown that the observed
strong radial magnetic field near the GC cannot be generated by the
most effective mechanism (producing the magnetic field) known so
far, e.g. the $\alpha$ dynamo and by all other mechanisms (producing
the magnetic field) proposed in recent years. In addition, the
observed powerful radial magnetic field at $r=0.12\rm{pc}$  from the
GC with magnetic field strength $B>$mG
 \citep{Eatough13} is almost the same as our prediction (at
$r=0.12\rm{pc}, B=(10-50)$mG based on the model of super-massive
object containing MMs at the GC.

The above three predictions (the first, third, and last one) are all
completely independent. These predictions are all consistent with
observations. These predictions are all consistent with
observations. These predictions cannot be accidental coincidence.
The discovery of powerful radial magnetic field near the GC  may
have the following two important physical significance: 1) the fact
that unusually strong radial magnetic fields are discovered near the
GC may be the convincing astrophysical evidence for the existence of
MMs. Thus our model for quasars and AGNs containing MMs is
reasonable. 2) The radiation originate from the direction of the GC
cannot be generated by the standard model of the black hole with its
accretion disk.

\subsection{On the MMs captured by stars and planets}

Stars and planets are formed from the gravitational collapse of
massive neutral hydrogen cloud. Since the interaction of MMs with
the neutral hydrogen cloud is very weak, there are very few MMs
accumulated at the stellar center following gravitational collapse.
However, stars and planets can capture MMs from interstellar space
during their life time after their birth. The number of the MMs
captured is directly proportional to the surface area and the age of
the stars and planets.  Making use of the number of MMs captured
from space (for stars and planets) and using the RC effect as the
energy source, we will mow consider a series of juggling
astrophysical phenomena including supernova explosion mechanisms of
curious types and $\gamma$-ray burst mechanisms.

\section{Supernovae Explosion Mechanism driven by MMs}
The masses for the progenitors SNII, SNIB, SNIC, SLSN are
respectively given by $(8-25) M_\odot, (30-60) M_\odot, (80-150)
M_\odot$, and $(200-1000) M_\odot$.

\subsection{The number of MMs captured by massive stars during their life time}

The time scale of the stars in the main sequence stage is about 90\%
of the life time for the stars. Due to hydrogen burning in the main
sequence stage of the stars the stars is relatively stable and the
stellar radius changes little during the main sequence stage.
However, the stellar radius during the red giant stage of the stars
will increases up to more than (10-100) times of the radius of the
main sequence stars at least , although the time scale of the stars
in the red giant stage is only about 10\% of the life time for the
stars. Thus, the total number of MMs captured from space by the
progenitor of supernova in its life time is mainly in its red giant
stage.

Making use of the radius and life time of the various types of
supernovae progenitors during their main sequence stage, we may
calculate the number of MMs captured during the main sequence life
time and multiplied. By multiplying the factor 10, we may estimate
the number of MMs captured by supernovae progenitors during their
life time (actually this is a lower limit). The MMs accumulated at
the central core of the various types of the supernovae are
originate from the from MMs captured in flight by their progenitors,
so their number is directly proportional to both the surface area
and the life time of the progenitors mainly during their red giant
stage. In the following we list the lower limit for the number of
MMs contained in the deep interior of the core for the various types
of supernovae (with typical masses)
\begin{equation}
\label{eq.13}
N_m=4\pi
R^2\phi_mt\approx1.0\times10^{31}(\frac{\phi_m^0}{10^{-2}\phi_m^{\rm{up}}})(\frac{R_{\rm{RG}}}{10^3R_\odot})^2(\frac{t_{\rm{RG}}}{10^6\rm{Yr}})
\end{equation}
where $R_{\rm{RG}}$ denotes the radius of the star in its red giant
stage and $t_{\rm{RG}}$ is the life time during the red giant stage
of the progenitor of supernova.

Because the MMs accumulated at the central core of the various types
of the supernovae are originate from the from MMs captured in flight
by their progenitors£¬so their number is directly proportional to
both the surface area and the life time of the progenitors during
their main sequence stage. These super-heavy MMs  must fall and
accumulated at the deep interiors of the central core. In the
central core the number of the MMs reach the Newtonian saturation
value such that $n_m(r)\approx\xi_s n_{\rm{B}}(r),
\xi_s\approx1.9\times10^{-32}(m_m/(10^9m_p))$. In this way, may
estimate the radius of the central core where the MMs accumulated as
follows
\begin{equation}
\label{eq.14}
 N_m\approx\frac{4\pi}{3}r_c^3\zeta_s\overline{n_{\rm{B}}}\approx1.3\times10^{24}(\frac{\phi_m^0}{10^{-2}\phi_m^{\rm{up}}})(\frac{\overline{n^c_{\rm{B}}}}{n_{nuc}})(\frac{r_c}{10^6\rm{cm}})^3
\end{equation}
where
\begin{equation}
\label{eq.15}
(\frac{r_c}{10^6\rm{cm}})\approx(\frac{\overline{n^c_{\rm{B}}}}{n_{nuc}})^{-1/3}(\frac{N_m}{1.3\times10^{24}})^{1/3}
\end{equation}

It is well known that the central density of massive stars is very
low before supernovae explosion. At least during the hydrogen
burning main sequence stage, from relations for the same mode of
stellar structure, more massive stars have lower central densities.
This kind of stellar structure will not change qualitatively during
the subsequent nuclear burning stages. The central density of the
Sun is $(50-100)$ $\rm{g/cm^3}$. From this it may be estimated that
the central density of massive stars are about 10 $\rm{g/cm^3}$
during their main sequence stage. In this way from Equ.(15), the
radius of the central core where the MMs accumulated may be
estimated be $r_c\approx(10^5-10^7)$ km, very much smaller than the
radius of the massive stars of the progenitor of the SN. During the
core collapse of the supernova, the MMs in the region within r
collapse together with the hot plasma toward the center of the star
by  the strong electromagnetic interaction of the MMs with the
plasma. However, during the gravitational collapse of the stellar
deep central core the MMs accumulated within the radius
$r_c\approx(10-10^3)$ km may interact strongly with the high
temperature plasma through the electromagnetic interaction

The MMs in the region within r  collapse together with the hot
plasma toward the center of the star. The condition for saturation
$n_m(r)\approx\xi n_{\rm{B}}(r)$ is still maintained during the
process of gravitational collapse. When the central density reaches
nuclear density during the collapse. We may use Equ.(10, 11) to
estimate the radius of the region for the accumulation MMs to be
only $r_c\approx(10-10^3)$ km . In the absence of the RC effect the
traditional theory for supernovae explosion is as follows. When core
collapse of massive stars reaches nuclear density£¬the
non-relativistic degenerate nucleon gas dominates the core pressure
(which is much larger than the relativistic and degenerate election
pressure), matters no longer continue to collapse inwards and a
strong outward rebound shock is formed. However, as we discussed in
Section 1.1, traditional theory cannot trigger supernova explosion.
On the other hand£¬in the presence of the RC effect, when the
central density approaches the nuclear density, only a very small
amount of the MMs (less than $10^7$ mole) accumulated in the central
core of radius $r_c\approx(10-10^3)$ km can trigger continuous
nucleon decay. And the resulting RC luminosity with the increasing
central density of the nucleons following the continue collapsing is
drastically increased to exceed the Eddington luminosity. The
corresponding radiation pressure far exceeds the non-relativistic
nucleon degenerate pressure. The strong radiation pressure will make
the entire star to expand outwards. In other words, the huge
radiation pressure due to the RC effect can trigger supernovae
explosion.

\subsection{The RC luminosity of various types of supernovae}

Using nucleon decay induced by MMs as the energy source, we will now
discuss the explosion mechanisms of the various types of supernovae
(SNII, SNIb, SNIc and and SLSN) and $\gamma$-ray bursts (including
long ray burst and short $\gamma$-ray burst).  As before, we adopt
the parameter value $\xi\approx50$. At the end of thermal nuclear
evolution of the super-massive stars (SNIc and the progenitor of
SLSN) the process of electron- positron pairs annihilation leads to
the unstable collapse of the central core of such stars. When the
baryon number density exceeds the nucleon density the MMs
accumulated in the deep interior of the stellar core quickly trigger
nucleon decay via the RC effect and the energy released is very high
with highest efficiency. We now substitute the number of MMs
accumulated at the central core of the various types of supernovae
from Equs. (8) into Equ.(12) to obtain
\begin{equation}
\label{eq.16}
 L_m\approx2.5\times10^{43}a(\frac{\xi}{10^2})(\frac{n_{\rm{B}}^c}{n_{nuc}})(\frac{T_c}{10^{11}\rm{K}})^{1/2}~~\rm{ergs/s}
\end{equation}
where
\begin{equation}
\label{eq.17}
a=(R_{\rm{RG}}/(10^3R_\odot))^2(t_{\rm{RG}}/(10^6\rm{Yr}),
\end{equation}
\begin{equation}
\label{eq.18}
\xi\equiv\sigma/(10^{-30}\rm{cm}^3)(\phi^0_m/(10^{-2}\phi^{\rm{up}}_m)).
\end{equation}
In Equ.(16) of the RC luminosity of supernova, parameter a is
determined by  both the radius, $R_{\rm{RG}}$, and the life time,
$t_{\rm{RG}}$, of the progenitor of the SN in the red giant stage.
The parameter $\xi$ is the product of five uncertain physical
quantities. Although the two factors, $ \sigma/(10^{-30}\rm{cm}^3),
(\phi^0_m/(10^{-2}\phi^{\rm{up}}_m)) $ are uncertain, but the
typical value of the parameter $\xi$ in Equ.(16) may be taken as
$\xi=100$ by comparing the heat flux for hot molten interior of the
earth with the same Equ.(16) in the unified model ( see Section
5.2).

\subsection{Estimate of the radius, main sequence life time and the parameter $a$ for the progenitor of SNII}

If radiation pressure is neglected, the upper half main sequence
masses are larger and the mass-radius relation is approximately
$R\propto M^{2/3}$. The corresponding mass-luminosity relation for
the upper half main sequence stars with larger masses are
$L/L_\odot\approx(M/M_\odot)^{3.5}$. The continuous hydrogen burning
($4^1\rm{H}\rightarrow^4$He) in the central region releases 26.73MeV
with efficiency $7.1\times10^{-3}$. The hydrogen abundance of
nascent stars is denoted by $X_0$ (where $X_0$ is roughly about 0.68
for the sun). When a fraction, f ($f\approx0.12$ ) , of the hydrogen
of the entire star has converted into helium, the star leaves the
main sequence. About 90\% of the life time of stars are in the main
sequence. The main sequence life time $t$ of a massive star with
initial main sequence mass M may be roughly written as
\begin{equation}
 \label{eq.19}
t_{\rm{MS}}=\frac{7.1\times10^{-3}fMc^2}{L}\approx1.1\times10^{10}(\frac{fX_0}{0.1})(\frac{M}{M_\odot})^{-2.5}~~\rm{Yr},
\end{equation}

The mass-radius relation and the mass-luminosity relation given
above can be approximately applied to the progenitor of SNII(Their
initial mass is less than 30 $M_\odot$). When the initial
main-sequence mass of a star is greater than 20$M_\odot$, the
fraction of the radiation pressure to the total pressure begins to
exceed ($25\%-30\%$) and radiation pressure cannot be neglected .The
radiation pressure increase very fast as the stellar mass increase.
The effect of the radiation pressure on the stellar radius cannot be
neglected. The stellar radius increase very fast as the stellar mass
increases and the stellar radius clearly grows to exceed the
previously mentioned limit for which the radiation pressure is
neglected ($R\propto M^{2/3}$). For the stars with still larger mass
($M>M_\odot$) (They are the progenitors of SNIb SNIC SLSN) , the
radiation pressure dominates the total pressure and the stellar
radius increases very fast as the stellar mass increases. For a
super-massive star ($M>100 M_\odot$), radiation pressure dominates
with an approximate mass-luminosity relation $L\propto M$
\citep{Shapiro83}. For these massive stars, the radiation pressure
increases very fast as the masses are increased. Stellar winds are
drastically strengthen and no clear mass luminosity relation exists.
The stellar radius increases rapidly with increasing mass. We take
the following crude approximation (hypothesis)
\begin{equation}
\frac{R_{\rm{MS}}}{R_\odot}=10(\frac{M}{20M_\odot})^{\beta}~~~~~~~~~~~(\beta\approx1.0-1.5),
 \label{eq.20}
\end{equation}
and we have
\begin{equation}
\label{eq.21}
\frac{R_{\rm{RG}}}{R_\odot}=10^3(\frac{M}{20M_\odot})^{\beta}~~~~~~~~~~~(\beta\approx1.0-1.5),
\end{equation}
which for the radius of the progenitor of the SN with the red giant
stage.

Main sequence life time for such super-massive stars can only be
roughly estimated
\begin{equation}
\label{eq.22}
 t_{\rm{MS}}= 10^7(\frac{M}{20M_\odot})^{-1}~~\rm{Yr}.
\end{equation}
The life time of the red giant stage of such super-massive stars may
be estimated as
\begin{equation}
\label{eq.23}
 t_{\rm{RG}}= 10^6(\frac{M}{20M_\odot})^{-1}~~\rm{Yr}.
\end{equation}

From Equ. (17), with the two Equs. (21, 23), we may crudely estimate
the parameter  $a$ by a relationship with the mass of the
supermassive stars, $a=(M/20M_\odot)^{2\beta-1}$. Substituting it
into (16), the RC luminosity catalyzed by the MMs accumulated at the
central core of the supernovae is obtained as following
\begin{eqnarray}
\label{eq.24}
 L_m\approx2.5\times10^{43}(\frac{M}{20M_\odot})^{2\beta-1}(\frac{\xi}{100})\nonumber\\
 \times(\frac{n_{\rm{B}}^c}{n_{nuc}})(\frac{T_c}{10^{11}\rm{K}})^{1/2}~~\rm{ergs/s}
\end{eqnarray}

The core collapsing process is close to the free collapsing process
(the inward collapsing velocity is about half of the free fall
velocity). For more massive stars, the free collapsing velocity is
faster and the more tightly compressed core, then leads to larger
$n_{\rm{B}}^c/n_{nuc}$. Using the relation between the radius and
mass of the degenerate compact stars  $R\propto M^{-1/3}$, we
speculate that the center density of collapsed core of the supernova
is approximately $n_{\rm{B}}^c/n_{nuc}\propto (M/20 M_\odot)^2$.
Substituting it in to the Eq.(24), we obtain
\begin{equation}
\label{eq.25}
 L_m^{\rm{peak}}\approx2.5\times10^{43}(\frac{M}{20M_\odot})^{2\beta+1}(\frac{\xi}{100})(\frac{T_c}{10^{11}\rm{K}})^{1/2}~~\rm{ergs/s}
\end{equation}

Thus we have
\begin{equation}
\label{eq.26}
 L_m^{\rm{peak}}\propto(\frac{M}{20M_\odot})^3~~~~~~~(\beta=1)
\end{equation}
\begin{equation}
\label{eq.27}
 L_m^{\rm{peak}}\propto(\frac{M}{20M_\odot})^4~~~~~~~(\beta=1.5)
\end{equation}

For the brightest known super luminous supernova (SLSN)
\citep{Dong16}, if the mass of the initial main sequence star for
the progenitor of the SLSN exceeds $(10^3-10^4) M_\odot$ or more,
then its radius during the main sequence stage is about $(10^4-10^5)
R_\odot$, but its main sequence lifetime is very short, only
$(10^4-10^5)$ years. Thus the parameter   defined by (16) may reach
( $10^6-10^7$). The RC luminosity of the energy release rate will
reach more than $(10^{49}-10^{50})\rm{ergs/s}$  when the core center
density of the supernova exceeds the nuclear density. This explains
the observed luminosity for the brightest super luminous supernova
so far.

When the central density of the collapsed core exceeds the nuclear
density, then the rates of energy release from the central core due
to the RC effect for supernovas of types SNII, SNIb, SNIc and the
brightest supernovae SLSN can all reach $10^{43} \rm{ergs/s}$ or
more. The huge radiation pressure drives the central core of the
stars to violently expand outward to form SN¢ò, SNIb , SNIc, and
SLSN. This is the effective mechanism for supernovae explosion
driven by the MMs. With increasing mass of the progenitor for the
series SNII, SNIb, SNIc and SLSN, the peak luminosity of the
supernova rapidly increases with $(M/20 M_\odot)^{(2\beta+1)}$. The
central density of the collapsed core of the supernova is
approximately proportional to the square of the mass of the
progenitor of the supernova.

\subsection{The concrete mechanism for supernovae explosion}

In order to explode the entire star violently, the RC luminosity
must much exceed the Eddington's luminosity. Therefore, we have
\begin{equation}
L\gg L_{Edd}=\frac{4\pi
cGM}{\kappa}\approx1.3\times10^{38}(\frac{\kappa}{0.4})^{-1}(\frac{M}{M_\odot})~~\rm{ergs/s},
 \label{eq.28}
\end{equation}
where $\kappa$ denotes opacity. At high temperatures Thomson
scattering dominates $\kappa\approx0.4\rm{g cm^2}$. When
condition£¨the inequality (28) is satisfied, the corresponding
radiation pressure is so huge that the stellar mantle, outer layer
and the stellar atmosphere of the entire star are all violently
ejected outwards. The highest speed of the drastic projection may
reach $(1-2)\times10^4\rm{km/s}$ . This is the observed scenario of
supernovae explosion. As for how large the ratio of
$L_m/L_{\rm{edd}}$ must be in order to achieve supernovae explosion,
detailed numerical simulation is really needed. Here we may roughly
estimate the ratio
\begin{eqnarray}
b=L_m^{\rm{peak}}/ L_{\rm{Edd}}\approx
2\times10^4(\frac{M}{20M_\odot})^{2\beta-1}(\frac{n_B^c}{n_{nuc}}),
 \label{eq.29}
\end{eqnarray}

Therefore, the value of b has already reached 400 not yet reach 1\%
of the nuclear density even for the initial main sequence mass with
20$M_\odot$. Thus the RC luminosity of the supernova must be much
greater than the Eddington's luminosity i.e. the condition Equ.(28)
is satisfied£¬the MMs accumulated at the central core of massive
stars can continuously induce nucleon decay leading to huge RC
luminosity and drive violent supernovae explosion.

We note that the Eddington luminosity increases with the masses of
the supernovae progenitors of the series SNII, SNIb, SNIc, SLSN
increase. In order to achieve supernovae explosion£¬the supernovae
core must he highly compressed to exceed nuclear density such that
the required RC luminosity is reached. Actually, during the
gravitational collapse of massive stars election capture in iron at
high density is very fast, both the resulting number of free
elections and the degenerate election pressure drop very quick and
the core collapse is close to free collapse (the inward collapsing
speed in half that of the free collapse).

More massive stars have larger free collapsing speed, higher density
of the stellar core, larger  $n_B/n_{nuc}$ ratio and greater RC
luminosity. Consequently, as the masses of the progenitors of the
series SNII, SNIb, SNIc, SLSN increase, the peak value of the
supernovae explosion becomes higher. The is basically consistent
with general consensus.

In the following, we will discuss in more detail how the huge RC
radiation pressure blow off the supernovae. In the region $r\leq
r_c$, the RC luminosity may be written as
\begin{eqnarray}
L_m(r)=\int_0^r4\pi r^2n_m(r)n_{\rm{B}}(r)\langle\sigma v\rangle m_{\rm{B}}c^2dr \nonumber\\
=\int_0^r4\pi r^2\zeta_sn_{\rm{B}}^2(r)\langle\sigma v\rangle
m_{\rm{B}}c^2dr
 \label{eq.30}
\end{eqnarray}

\begin{eqnarray}
L_m(r)=(4\pi/3) r^3\zeta_s\langle\sigma v\rangle m_{\rm{B}}c^2(\overline{n_{\rm{B}}^c})^2\nonumber\\
\approx0.7\times10^{40}(\frac{\overline{n_{\rm{B}}^c}}{n_{{\rm{nuc}}}})^2(\frac{r}{10^6\rm{cm}})^3(\frac{\sigma}{10^{-30}\rm{cm}^2})(\frac{T}{10^{11}\rm{K}})^{1/2}
 \label{eq.31}
\end{eqnarray}
where $\overline{n_{\rm{B}}^c}$  denotes the average nucleon number
density of the central core in which the MMs are accumulated. The
temperature of the central core is $10^{11}$K  with corresponding
nucleon thermal velocity $5\times10^{9}\rm{cm/s}$. Since all the MMs
are concentrated within $r\leq r_c$, there are no magnetic monopoles
outside $r_c$, therefore the RC luminosity for $r> r_c$  is
constant, namely
\begin{eqnarray}
L_m(r)\approx 0.7\times10^{40}(\frac{\overline{n_{\rm{B}}^c}}{n_{nuc}})^2(\frac{r_c}{10^6\rm{cm}})^3\nonumber\\
~~~~\times(\frac{\sigma}{10^{30}\rm{cm}^2})(\frac{T}{10^{11}\rm{K}})^{1/2}~~~~~~~~(r>r_c)
 \label{eq.32}
\end{eqnarray}

The huge radiation pressure generated  by the RC luminosity may be
determined by
\begin{equation}
 -\frac{dP_r(r)}{dr}=\frac{\kappa\rho(r)L_m(r)}{4\pi r^2c},
 \label{eq.33}
\end{equation}
Substituting (31) and (32) into (33), we obtain
\begin{equation}
 -\frac{dP_r(r)}{dr}=2.0\times10^{42}(\frac{\xi}{100})(\frac{\kappa}{0.4})(\frac{\overline{n_{\rm{B}}^c}}{n_{nuc}})^3(\frac{r_c}{10^6\rm{cm}})(\frac{r}{r_c}),
 \label{eq.34}
\end{equation}
\begin{eqnarray}
 -\frac{dP_r(r)}{dr}=2.0\times10^{45}(\frac{\xi}{100})(\frac{\kappa}{0.4})(\frac{n_{\rm{B}}^c}{n_{nuc}})\nonumber\\
 \times(\frac{\overline{n_{\rm{B}}^c}}{n_{{\rm{nuc}}}})^2(\frac{r_c}{10^6\rm{cm}})^2(\frac{r}{r_c})^{-2},
 \label{eq.35}
\end{eqnarray}
where the conditions for Equs. (34, 35) are corresponding to $r\leq
r_c$ and $r>r_c$.

In the collapsing central region the gas pressure is dominated by
the non-relativistic neutron degenerate pressure with equation of
state
\begin{equation}
P_g(r)\approx K
\rho^{5/3}~~~~~~~~~~(K\approx5.4\times10^9~~\rm{c.g.s}),
 \label{eq.36}
\end{equation}
and the gas pressure gradient is
\begin{equation}
 -(\frac{dP_g(r)}{dr})\approx 1.1\times10^{34}(\frac{\rho(r)}{\rho_{{\rm{nuc}}}})^{2/3}(-\frac{d\rho(r)/d\rho_{{\rm{nuc}}}}{dr}),
 \label{eq.37}
\end{equation}

In stellar interior even if we take the high density gradient
$d\rho(r)/d\rho_{\rm{nuc}}\approx10^5$ km (a fast decrease of
nuclear density to $10^8\rm{g/cm^3}$ in a distance of 10 km) the
radiation pressure generated by the RC effect in the supernovae core
$<10^3$ km is far larger than the non-relativistic neutron
degenerate pressure. This means that we can neglect the
non-relativistic neutron degenerate pressure in our discussion, i.e.
the total pressure of stellar interior  $P(r)\approx P_r(r)$.
According to the theory and dynamic equation of stars, we have
\begin{equation}
 \rho(\frac{d^2r}{dt^2})= -(\frac{dP_r}{dr})-(\frac{GM(r)}{r^2})\rho(r),
 \label{eq.38}
\end{equation}
when
\begin{equation}
  -(\frac{dP_r}{dr})\gg(\frac{GM(r)}{r^2})\rho(r),
 \label{eq.39}
\end{equation}

The huge radiation pressure will make the stellar matter within r to
violently ejected outwards. We will introduce the parameter $b^{'}$
to depict the necessary condition that the radiation pressure is
sufficient to trigger supernovae explosion
\begin{equation}
  -(\frac{dP_r}{dr})\geq b^{'}(\frac{GM(r)}{r^2})\rho(r)~~~~~~~(b^{'}\approx10-10^2),
 \label{eq.40}
\end{equation}
where we assume that $b^{'}$ in Equ.(40) is roughly equal to the
value b in Equ.(29)).

When the density of supernovae core satisfied condition(the
inequality (29), the inequality (39) may also be satisfied and the
series of supernovae SNII, SNIb, SNIc, and SLSN explosions can
succeed. Our important conclusions are as follows, the necessary and
sufficient condition for the magnetic monopoles induced nucleon
decay and the resulting RC luminosity and radiation pressure
sufficiently strong to violently eject the stellar matter is that
the density of the collapsing core must reach near the nuclear
density and above.

The density of the collapsing core increases as the initial main
sequence mass of the supernovae progenitor series SNII, SNIb, SNIc,
and SLSN increase. From Equ.(16) and Equ. (25) we note also that the
RC luminosity and the peak luminosity increase very fast as the
initial main sequence mass increases.

Our model is different from the standard model SNII for supernovae
explosion. In Section 1, we already mentioned the reasons for the
unsuccessful SNII model. Firstly, the instantaneous explosion
mechanism for supernovae explosion fails due to the energy of the
outgoing rebound shock being exhausted before the matters in the
outer layers are entirely and completely destroyed by the
$\gamma$-ray photons (from the rebound shock). In our theory, the
MMs accumulated in the supernovae inner core continuously trigger
nucleon decay with huge energy release. When the density of the
collapsing core reaches or above nuclear density, the RC luminosity
may reach $10^{41}-10^{42}\rm{ergs/s}$  or above. This luminosity is
generated continuously and it provides powerful radiation flux until
supernovae explosion is achieved. This RC luminosity can only be
weaken after supernovae explosion when the SNII theory. The energy
of the outgoing shock is completely exhausted before the outer iron
core breaks up. Clearly, similar problem cannot happen in our
theory.

Secondly, our approach is also different from the neutrino delayed
explosion mechanism proposed by \citet{Wilson88} . supernovae
explosion cannot be achieved because the interaction between the
neutrino flux and matter is too weak to break the outer atmosphere
and supernovae explosion \citep{Buras03, Lie04}. In our theory the
successful supernovae explosion is achieved through the continuous
generation of the RC luminosity and the huge radiation pressure. MMs
triggered nucleon decay play the key role. Of course, in our
explosion scenario, the transformation of the gravitational
potential energy of $(10^{52}-10^{53})$ ergs convert into the
thermal energy. The powerful neutrino flux of energy $10^{52}$ ergs
and above generated by the conversion of (u, d) and (u, d, s) quarks
in the high temperature environment of the nascent neutron star are
the same as that of the standard SNII model\citep{Dai95}.

\subsection{Weak explosion or dark explosion of supernova}

If the RC luminosity is not much higher than the Eddington's
luminosity of the star during the collapse of the central core, then
the resulting explosion is rather weak and may be referred to as
dark explosion. The supernovae remnant Cas A (it is only 3.4 Kpc
away from the Earth), for instance, corresponds supernovae with
maximum visual luminosity of $5^m$, which exploded in 1680. The
remnant neutron star of this explosion was discovered in 1999. It is
very likely that this is a concrete example of weak explosion.

By researches via the observation of the Chandra X-ray satellite,
recently, the interstellar nebula G$1.9 + 0.3$ is considered as a
supernova remnant exploded about 110 years age, but it has never
been reported. It may be the dark explosion.These dark supernova may
be formed from the direct collapse of the white dwarfs. Their
density is not very high during supernova explosion, so the RC
luminosity is only slightly greater than Eddington luminosity and
they are weak explosion.

More recently, NASA reported on 28th May, 2017 by blasting news in
the network. A star N6946-BH1 suddenly disappears from astronomical
observations in 2015 (It also totally disappears even though from
observations by both Hubble Space Telescope and Spitzer Space
Telescope), although it is rather luminous in 2007 and its
brightness begun to strengthen in 2009. People guess that it has
directly collapsed to a black hole according to the popular idea.
But it is an typical example for the weak explosion in our model.

\subsection{The remnant neutron stars after supernovae explosion}

MMs induce nucleon decay in the supernovae core with huge energy
release. The resulting super-strong radiation pressure resists the
gravitation collapse of the supernovae core and the central core
cannot continue to collapse. The density of the core cannot approach
infinity. Instead, the matters in the supernovae core are driven
outwards by the huge radiation pressure and the density of the
supernovae core drops quickly.

When the RC luminosity is strong the velocity of the large amounts
of matter projected outwards exceeds the escape velocity. Thus,
during supernovae explosion, the mantle, outer layer and the outer
atmosphere are all ejected far away from the star. But, when the RC
luminosity and the corresponding radiation pressure are lowered, the
matters in the stellar interior with ejection velocity less than the
escape velocity and especially the matters in the deep supernovae
core begin to fall towards the center. But the resulting
gravitational collapse will not lead to infinite density to form
black hole. This is because there are always a few MMs still remain
in the deep interiors of the supernovae core even though large
amounts of the MMs are quickly ejected outward outside the star. The
remaining MMs interact strongly with the high temperature plasma in
the supernovae core through electromagnetic interaction. These
remnant super heavy MMs in the innermost supernovae core can still
continuously to trigger nucleon decay to generate the RC luminosity
far lower than the Eddington luminosity of the remnant star.
Finally, according to Eq.(28), some kind of stable dynamical
equilibrium may be reached
\begin{eqnarray}
 L_{Edd}&=&1.3\times10^{38}(\frac{\kappa}{0.4})^{-1}(\frac{M}{M_{\bigodot}})
 \nonumber\\
 &&\gg L_m=\frac{4\pi}{3}r_c^3\zeta_s(n_{\rm{B}}^c)^2\langle\sigma v\rangle m_{\rm{B}}c^{2},
 \label{eq.41}
\end{eqnarray}

We may use Eq.(14) and Eq.(41) to estimate  of the central core
\begin{equation}
 \frac{n_{\rm{B}}^c}{n_{{\rm{nuc}}}}\ll 10^{-2}\eta(\frac{M}{M_\odot}),
 \label{eq.42}
\end{equation}
where
\begin{equation}
 \eta=[(\frac{m_{m}}{10^9m_p})(\frac{\sigma}{10^{-30}\rm{cm}^2})(\frac{N_m}{10^{24}})]^{-1},
 \label{eq.43}
\end{equation}

Only a few magnetic monopoles remains in the core of the nascent
neutron star ($N_m$ is much less than $10^{24}$) because most of the
magnetic monopoles in the collapsed core of the supernova are thrown
out with the plasma by the strong electro-magnetic interaction. From
this we can estimate the average matter density in the innermost
core at dynamical equilibrium to be only
$\rho_c\ll10^{-2}\rho_{\rm{nuc}}$. Outside the innermost core with
radius $r_c$, the mass of the matter contained increases very fast
and $L_{\rm{Edd}}$ also drastically increase. But the RC luminosity
$L_{m}$ original from the innermost core can no longer increase, so
$L_m\ll L_{edd}$, matter in the stellar interior can no longer be
driven outwards. Actually, RC luminosity $L_{m}$ decreases very fast
as the density of the supernovae core drastically decreases. From
Eq.(35), we note that the radiation pressure decreases very fast as
the ratio $\overline{n_{\rm{B}}^c}/n_{\rm{nuc}}$ decreases. It is
seen from Eq.(42) that if $\overline{n_{\rm{B}}^c}/n_{\rm{nuc}}\ll
10^{-2}$, the corresponding RC luminosity greatly decreases such
that the matter outside the innermost core containing MMs (with
radius $r_c$) can no longer be driven out. At this time the strong
gravitation due to the matter in the thick outer layer can compress
the matter inside the star to nuclear density to form neutron stars
or super-massive neutron stars. Using MMs induce nucleon decay as
the energy source, neutron stars cannot collapse to form black holes
no matter how massive they are.

The remnants after supernovae explosion are stellar objects similar
to neutron stars. The density in the innermost core is not high.
Because the remnant MMs can continuously induce nucleon decay to
provide energy source and huge radiation pressure so that this
stellar object cannot collapse to form a black hole. Other important
prediction from our analysis is that there is no upper limit for the
mass of the neutron stars. At least, there is no generally accepted
upper limit of 25$M_\odot$. A massive neutron star origins from the
supernova with a super massive progenitor. It is well known that the
probability of the birth of super massive stars is very small
according to the initial mass distribution function of Salpeter. Up
to now, the number of neutron stars that their masses have been
measured or (have been estimated) in close binaries is less than 20.
The above result without the upper limit of 25$M_\odot$ is not
inconsistence with  astronomical observations.

We would like to emphasize that the only difference between the
neutron stars formed from the standard SNII model and our model is
the innermost structure with MMs that we proposed. As for SNIb,
SNIC, and SLSN with their much more massive progenitors would not
collapse to form black holes with infinite density. Instead, they
form more massive neutron stars with remnant MMs accumulated in the
deep interior of the supernovae core. Since the long $\gamma$-ray
burst originated from supernovae SNIc, so that our model of
supernovae explosion driven by MMs also apply.

\section{The other related puzzle of our model of supernova driven by magnetic monopoles}

\subsection{The problem of the heat source of the core molten state in the Earth
interior}
\subsubsection{The puzzle of a hot molten state for the
core of the Earth}

It is well known from the eruption of the volcano and the hot magma
that the core of the earth is in a hot molten state, The core of the
Earth is very hot, but the shell temperature is relatively lower.
For instance, the temperature in the surface of the earth is 300K.
This state of molten core can't be caused by absorbing energy from
the solar radiation. Thus, it must contain another energy sources in
the interior of the earth. A popular idea on this question is that
the huge thermal energy in the Earth's core origins from collapsed
core in the formation process (for an example, see the book from
Wikimedia). However, you may see that this idea is questionable by
following discussion.

The total thermal energy from the interior of the Earth is given by
\begin{equation}
\label{eq.44}
  E_{k}\approx \frac{4\pi}{3}R_c^3 \overline{\rho_c}N_A k \overline{T}/\overline{\mu},
\end{equation}
where the core radius of the earth is $R_c\approx2\times10^3$Km. The
core density of the earth is
$\overline{\rho_c}\approx13\rm{g/cm^3}$. The core temperature of the
earth is $\overline{T_c}\approx6\times10^3$K. The average molecular
weight in the core of the Earth is $\overline{\mu}\approx30$. Thus,
we have $ E_{k}\approx 0.7\times10^{37}$ergs.

The total heat flow from earth's interior to surface crust is about
47 TW (i.e. $J_r\approx4.7\times10^{20}\rm{ergs/s}$)
\citep{Davies10}. The time scale of outward transport thermal
energy, in a case without energy source, from the interior of the
earth is given by
\begin{equation}
\label{eq.45}
  t_{th}=\frac{E_T}{J_r}\approx 1.5\times10^{16}s\approx0.5\times10^9\rm{Yr},
\end{equation}

This time scale is much shorter than the age of the earth. That is,
it is impossible that  the huge thermal energy in the Earth's core
is left over from the birth of the Earth. Therefore, we conclude
that there must be some heat source in the earth core. We first
discuss that if the radioactive elements in the interior of the
earth may provide enough energy in the time scale $t_{th}$. The heat
energy is mainly produced from the three most important radioactive
elements. The heat release rate by the three naturally radioactive
elements are $0.78 \rm{cal(yr g(U))}^{-1}$, $0.20 \rm{cal(yr
g(Th)})^{-1}$, $2.6\times10^{-6}\rm{cal(yr g(K))}^{-1}$ for the
uranium series, the thorium series, and the actinium series,
respectively \citep{Allen56}. As an example, we may discuss the most
important radioactive elements Uranium. According to some parameters
of the abundance of atomic number
$n(\rm{U})/n(\rm{H})\approx10^{-12}$, the mass abundance
$m(\rm{U})/m(\rm{H})\approx10^{-10}$ , and the mass of the earth
$6\times10^{27}$gm, we can know the total mass of radioactive
elements Uranium are about $6\times10^{17}$gm. Thus, the total heat
release of all radioactive elements Uranium from the interior of the
earth within $0.5\times10^{9}$Yr, is given by
$E_{\rm{U}}\approx10^{34}\rm{ergs}\approx1.5\times10^{-3}E_{\rm{T}}$

Synthesizes the above analysis, our conclusion has two. One is that
radioactive energy far cannot provide the source of energy of the
earth core melt state. Another is that the thermonuclear reaction
cannot be ignited due to the lower temperature of the Earth
interior(i.e. $6\times10^3$K ). Thus, it is necessary for us to find
new energy source for the melt state of the Earth core.

\subsubsection{The puzzle of a hot molten state for the
core of the Earth}

MMs accumulated in the central core of the earth and the RC
luminosity generated by nucleon decay induced by the MMs may be
computed from Eq.(9). The temperature of the earth at its center is
about $T\approx6\times10^3$K. According to our model for stellar
objects containing MMs, the RC luminosity generated by the number of
MMs that are captured by the Earth from interstellar space since its
birth may be estimated from Equ.(16)
\begin{eqnarray}
L_m&\approx3.0\times10^{18}(\frac{\sigma}{10^{-30}\rm{cm}^2})(\frac{\phi_m^0}{10^{-2}\phi_m^{\rm{up}}})\nonumber\\
&=3.0\times10^{18}\xi~\rm{ergs/s},
 \label{eq:46}
\end{eqnarray}
where
$\xi=(\frac{\sigma}{10^{-30}\rm{cm}^2})(\frac{\phi_m^0}{10^{-2}\phi_m^{\rm{up}}})$,
Compared with $J_r=4.7\times10^{20}\rm{ergs/s}$, if the outward heat
flow from the Earth's interior is provided by the RC effect we must
choose the parameter $\xi\approx100$. In other works, starting from
the actual date of the hot molten interior of the Earth, we are able
to determine the parameter $\xi\approx100$. In sections 4, we will
use this value of $\xi$ to determine the RC luminosity of supernova
explosion.

By the way, we may estimate the number of MMs captured by Jupiter
after its birth (the age of Jupiter is about 450 million years )
\begin{equation}
\label{eq.47}
N_m\approx5.0\times10^{19}(\frac{\zeta_m^0}{\zeta_s})(\frac{n_{\rm{B}}^0}{1\rm{cm}^{-3}})(\frac{\nu_m}{10^{-4}c})
\end{equation}

There are also the energy source problems for Jupiter, the
corresponding energy production rate is
$L_m\approx10^{21}\rm{ergs/s}$. Whether this is correct or not may
be tested by future study. Whether this is well or not may be tested
by future study.

\subsection{On Cooling of white dwarfs}
\subsubsection{The puzzle of White Dwarf cooling}

The effective temperatures of most of the white dwarfs (hereafter
WDs) are in the range of $5.5\times10^3-4\times10^4$K, only few WDs
have effective temperatures outside this range. Their spectral types
corresponding to these white dwarfs are from O to K.

But individual ones is for M types \citep{Zhu03}. Why are the
spectral types for most of WDs above types A (i.e. O, B, A) , but
only are F, G, K for few WDs, and few of WDs have the spectral types
of later M, and N, whose surface temperature are less than
$3\times10^3$K.

The temperature in the interiors of WDs is $10^6$K with total
thermal energy less than $10^{47}$ergs. The radius of white dwarf
star is about $10^4$ km with surface temperature $T
\approx5\times10^3-4\times10^4$K . Since most of WDs have surface
temperature $1\times10^4$K , we therefore adept the typical surface
temperatures of WDs as $1\times10^4$ K. The radiation luminosity of
WDs is roughly $L_{\rm{rad}}\approx10^{31}\rm{ergs/s}$, so the
typical cooling time scale of WDs is about
$10^6s\approx3.3\times10^8$Yr. In the other words, WDs. Should cool
down within 400 million years, then why no late type M and type late
N WDs with surface temperatures less than 3000K ever been observed?
What are the heat sources of these late type WDs? Astronomers did
not discuss this problem. This is because no other physical process
could provide such heat source. Adopting a model of celestial
objects containing the MMs and taking the RC effect into account as
an energy source(the MMs may catalyze nucleon decaying), we solve
the explosion mechanism in this paper for all kinds of supernova
including  $\gamma$ray burst and the energy sources in the core of
the earth and White Dwarfs.

\subsubsection{The RC luminosity in White Dwrafs interiors}

The MMs captured by WDs (including progenitor stars) from the
universe (accumulated at the stellar core) during their life time
and using the RC effect as the energy source may naturally explain
the existence of internal energy source in white dwarfs so that
white dwarf cooling can be stopped. According to the standard theory
of stellar evolution the WDs observed now originated from pre-main
sequence stars with masses $(2-8)$. After hydrogen and helium
burning stage these stars with low and middle mass lost their outer
layer through the AGB star stage, then their central cores become
white dwarfs.

We have shown in Section 3.3 that the total number of MMs captured
from space by the progenitor of the white dwarfs is about 10 times
of one calculated by Eq.(5) for this progenitor during its main
sequence stage with a radius about 2R? Thus from Eq.(5), we may
estimate the total number of MMs captured from the space by the
progenitor of white dwarfs during their life time to be
\begin{equation}
\label{eq.48}
N_m\approx1.0\times10^{28}(\frac{\phi_m^0}{\phi_m^{\rm{up}}})(\frac{R_{\rm{RG}}}{R_\odot})^2(\frac{t_{\rm{RG}}}{10^8\rm{Yr}})
\end{equation}
where $R_{\rm{RG}}$ is the radius of the progenitor of the write
dwarf in the red giant stage. Its typical value is taken as
$100R_{\bigodot}$? and the time scale of their red giant stars with
mass greater than 2m? is taken about $10^7$ years.

These captured super-massive MMs accumulated at the deep interior of
the stellar core and the resulting RC luminosity is again given by
Eq.(4). During the hydrogen burning main sequence stage, helium
burning red giant stage and the AGB stage for their progenitors, of
the white dwarfs, the stellar core density is respectively
$\rho_c=10, 10^3, 10^6\rm{g/cm^3}$ or $n_{\rm{B}}^c=10^{25},
10^{27}, 10^{30} \rm{cm}^3$, and the central temperature is
respectively $T_c=3\times10^7, 2\times10^8, 10^8$K, with
corresponding thermal speed, $\overline{\nu_T}\approx10^8\rm{cm/s}$
and luminosity $L_m\approx5\times10^{26}\xi, 5\times10^{28}\xi,
5\times10^{31}\xi~~\rm{ergs/s}$.

These luminosity is far lower than the corresponding luminosity
provided by thermal nuclear reaction during stellar evolution (where
$\xi\approx50$). After the evolutionary stage of white dwarfs,
however, thermal nuclear burning in the interior of white dwarfs
$T_c\approx(1-3)\times10^6$K stopped and there are no more energy
sources. The central temperature of white dwarfs   with
corresponding thermal velocity
$\overline{\nu_T}\approx10^7\rm{cm/s}$ . The typical value of the
central density for white dwarfs is $\rho_c=10^7\rm{g/cm^3}$ or
$n_{\rm{B}}^c=10^{30} \rm{cm}^3$. Then we obtain from Eq.(16)
\begin{equation}
\label{eq.49}
L_m\approx1.2\times10^{33}\xi(\frac{R_{\rm{RG}}}{R_\odot})^2(\frac{t_{\rm{RG}}}{10^8\rm{Yr}})
\end{equation}
where $R_{\rm{RG}}$ is the radius of the progenitor of white dwarfs
(i.e., the main sequence star) $t_{\rm{RG}}$ is the life time of the
main sequence star. If we use MMs induced nucleon decay as the
energy source to reach the white dwarfs luminosity
$L_{\rm{rad}}\approx10^{31}\rm{ergs/s}$, we are required to choose
\begin{equation}
\label{eq.50}
\xi(\frac{\nu_m}{10^{-4}c})(\frac{R_{\rm{RG}}}{R_\odot})^2(\frac{t_{\rm{RG}}}{10^8\rm{Yr}})\approx2
\end{equation}

If we also use the RC effect to provided the energy source for the
molten hot core of the earth, we must choose $\xi\approx50$ as we
did before consequently
\begin{equation}
\label{eq.51}
(\frac{\nu_m}{10^{-4}c})(\frac{R_{\rm{RG}}}{R_\odot})^2(\frac{t_{\rm{RG}}}{10^8\rm{Yr}})\approx0.04
\end{equation}

This is a constraint on the radius and life time of the progenitor
of white dwarfs (i.e., the constraint on the main sequence star).
The white dwarfs luminosity are really in the range
$10^{29}-10^{35}\rm{ergs/s}$, If we also use the RC effect to
provide the energy source for the molten hot core of the earth, we
must choose $\xi=100$. Thus we may get a constraint on the radius
and life time of the progenitor of white dwarfs during the red giant
stage
\begin{equation}
\label{eq.52}
(\frac{R_{\rm{RG}}}{R_\odot})^2(\frac{t_{\rm{RG}}}{10^7\rm{Yr}})\approx(10^{-4}-10)
\end{equation}

Equ.(40) is a reasonable constraint. We note that the observed date
for both molten state of the earth's core and the white dwarfs
luminosity are accurate and reliable. Therefore the value of $\xi$
that we derived and the constraint condition (40) are very
reasonable.

\subsection{The problem for the possible association of gravitational
waves and the short Gamma-ray burst}

Recently, the study of the astrophysics of gravitational waves has
attracted considerable attention. The international high frequency
gravitational Wave Detector project LISA (for the detection of
gravitational waves generated by the collision and merge of two
massive AGNs with super-massive black holes) is now in active
planning for space detection. The collision and merge of two smaller
compact stellar objects (black hole, neutron star and white dwarf)
may also produce gravitational waves. To this goal, the project for
ground based observations called Advanced LIGO has already begin to
accumulated large amounts of detected data LIGO already announced
the detection of gravitational waves GW150914 in February 12, 2016
 \citep{Abbott16}. Using the model of the collision of two black
holes with masses $36^{+5}_{-4}M_\odot$ and $29^{+4}_{-4}M_\odot$
respectively, the LIGO astronomers attempted to fit the formation of
a final black hole with mass $62^{+4}_{-4}M_\odot$ after the
generation of gravitational waves. The total energy released by the
gravitational waves is approximately $3^{+0.5}_{-0.5}M_\odot$.

A few days later (i.e., February, 16, 2016), the  $\gamma$-ray burst
detecting group, Fermi GBM in the U.S published a paper
\citep{Connaughton16} They reported that a weak and short
$\gamma$-ray burst (lasting one second) was almost simultaneously
detected (0.4 second later) in the direction of the gravitational
wave. From the fact that the non-thermal luminosity is
$10^{49}\rm{ergs/s}$  in the wave zone (1KeV-10MeV), they asserted
that this  $\gamma$-ray burst is associated with the LIGO
gravitational wave event GW150914.

But \citet{Lyutikov16} suggested a different view point concerning
the association of the gravitational wave event with the
$\gamma$-ray burst event. First, the $\gamma$-ray burst associated
with the gravitational wave event was not detected by another X-ray
space detector INTERGRAL. (However, the sensitivity in the 50KeV
energy range of the INTERGRAL X-ray space detector is not enough).
The author made a detailed analysis about all the possible
electromagnetic radiation mechanisms in the environment of a black
hole with accreting plasma and concluded that the observed
luminosity of $10^{49}\rm{ergs/s}$ is several order of magnitudes
higher than that can be provided by all the possible mechanisms
under consideration. From this he concluded that the $\gamma$-ray
burst event observed by Fermi GBM observations is not associated
with the gravitational wave event GW150914. The author believes that
these are two mutually independent coincident events in space and
time.

Whether or not the gravitational wave event is associated with the
$\gamma$-ray burst event? If the  $\gamma$-ray burst event observed
by Fermi GBM is real, then how to explain the association of the
gravitational wave event with the  $\gamma$-ray burst event? In his
exotic theoretical model, Zhang quickly gives the answer
\citep{Zhang16a, Zhang16b}. If one of the two merging black holes
carries enough and large amounts of electric charges (reaching
$10^{-4}$ of the critical charge $Q_c$) $Q\approx10^{-4}Q_c$ and
$Q=1.0\times10^{31}(M/10M_\odot)$ e.s.u, then not only gravitational
waves can be generated but also triggering the production of short
$\gamma$-ray burst and fast radio burst. \citet{Zhang16a, Zhang16b}
therefore concluded that the $\gamma$-ray burst event observed by
the Fermi GBM observations is associated with the LIGO gravitational
wave event GW150914.

Zhang also proposed that the gravitational wave event,  the short
$\gamma$-ray burst event and the fast radio burst event are somehow
associated with each other (see \citet{Zhang16a, Zhang16b}). The key
problem is that the charged black hole must carry large amounts of
charges. No body has never discussed how such highly charged black
holes are formed. The highest saturation value for the charges
carried by a Newtonian stellar object is roughly only about
$5.2\times10^{12}(M/10M_\odot)$ e.s.u.

The Newtonian saturation value of electric charges for a charged
celestial body is defined as follows. If a celestial body carries a
positive charge and when the Coulomb electrostatic repulsive force
acting on a proton is equal to the Newtonian gravity of the proton,
the electric charge of the celestial body is called as the Newtonian
charged saturation of protons. If the celestial body carries
negative charges, the Newtonian charged saturation of electrons is
about $1/1840$ of one of protons for the celestial body with
positive charge.

The critical charge of RN black hole is that when the electric
quantity of the black hole reaches the value, the visual interface
of the black hole disappears. If massive stellar objects really
carry huge electric charges before they gravitationally collapsed to
charged black holes, it may lead to unthinkable spectacular
phenomena due to electrical polarization effects. It is difficult to
believe that highly charged massive black holes really exist.

In our theoretical framework these two events are maturely
associative. Short $\gamma$-ray bursts are generally believed to
originate from the collision of neutron stars. The remnant neutron
stars after supernovae explosion may also contain some magnetic
monopoles in remnant neutron stars are evidently less than that in
the progenitor stars. Free MMs can also be captured from space after
the formation of neutron stars.

In our model of stellar objects containing MMs, there is a strong
radial magnetic field. Hence, during the collision and merge of the
two compact stellar objects (with same or apposite polarity) both
gravitational and electromagnetic interaction can participate
leading to radial attraction and repulsion. In the general frame
work of the two different models (the standard black hole model and
our model of super-massive stellar objects containing MMs) it is
expected that the wave forms of the gravitational waves generated by
the collision and merge of two AGNs (our model) or the collision and
merge of two compact objects (black hole, neutron star, white dwarf)
(the standard model) are rather different.

The collision and merge of two black holes can only generate
gravitational waves. But no $\gamma$-ray burst can be simultaneously
generated. However, the collision and merge of two compact and
massive stellar objects containing MMs (with central density far
exceed nuclear density reaching $(10^3-10^5)\rho_{nuc}$ or more the
resulting RC luminosity may produce $10^{49} \rm{ergs/s}$ and
simultaneously generate short $\gamma$-ray burst and large number of
charged particles via nucleon decay induced by the MMs. The radial
magnetic field from the MMs may in turn generate electric field in a
spinning stellar object to accelerate the charged particles. The
radial magnetic field lines may be distorted. Radio waves may be
generated when charged particles are ejected along curved magnetic
field lines. It is possible that the production of fast radio burst
(FRB) follows. From the above scenario, the short  $\gamma$-ray
burst occurred within a very short time interval after the
appearance of the gravitational waves. If the two events observed
are reliable, it seems advantageous to the validity of our model and
lend support to the key roles played by the MMs.

Of course, similar to the weak explosion and the dark explosion of
supernova, the short  $\gamma$-ray burst may not be observed, if its
RC luminosity is not much more than the Eddington's luminosity. The
second weaker LIGO gravitational wave event GW170104 without short -
ray burst may correspond to this situation in our model.

We will endeavor to further study and propose that if the
association of  $\gamma$-ray bursts with the production of
gravitational waves is observed again, the association of the two
events are real and it may be regarded as the observational evidence
for the existence of MMs.


\section{Conclusions and outlooks}

The series of astrophysical phenomena discussed in our paper has
been considered by astronomers and scientists as important but
puzzling problems in the recent half century. Especially the
explosion mechanisms for the various types of supernovae are the
central topics to study in the recent half century and no convincing
solutions to these puzzles has ever been found. Making use of the
two ideas of the spatial flux of magnetic monopoles and nucleon
decay induced by MMs (RC effect) as well as the theoretical formulae
in our papers \citep{Peng85a, Peng85b}, we are able to explain the
thermal energy source at the earth's central core, the heat source
needed in white dwarf interiors and the explosion mechanisms for
SNII, SNIb, SNIc, SLSN supernovae explosion and Gamma-ray bursts. We
would like to emphasize that in our model we just take the
parameter, $\xi$ (in Eq. (18)), as determined from the Earth's
thermal flux.

It seems that our unified treatment of these puzzling issues has no
doubt convinced us that the idea of MMs and the RC effect are
reasonable and powerful tools for future investigation. We believe
and we hope that our suggestions can attract considerable attention
and leading to another round of interest to study MMs .

\acknowledgments

We would like to thank the anonymous referee for carefully reading
the manuscript and providing some constructive suggestions which are
very helpful to improve this manuscript. One of us, C. K. Chou would
like to thank professor J. L. Han for his hospitality and
encouragement. This work was supported in part by the National
Natural Science Foundation of China under grants 11565020, 10773005,
and the Counterpart Foundation of Sanya under grant 2016PT43, the
Special Foundation of Science and Technology Cooperation for
Advanced Academy and Regional of Sanya under grant 2016YD28, the
Scientific Research Starting Foundation for 515 Talented Project of
Hainan Tropical Ocean University under grant RHDRC201701, and the
Natural Science Foundation of Hainan province under grant 114012.

\end{document}